\documentclass[10pt]{article}
\usepackage{amssymb}
\usepackage{amsmath}
\usepackage{graphicx}
\usepackage{fancyhdr}
\usepackage{cite}

\begin{document}

\newcommand{\bin}[2]{\left(\begin{array}{c} \!\!#1\!\! \\  \!\!#2\!\! \end{array}\right)}
\newcommand{\binc}[2]{\left[\begin{array}{c} \!\!#1\!\! \\  \!\!#2\!\! \end{array}\right]}
\newcommand{\threej}[6]{\left(\begin{array}{ccc}#1 & #2 & #3 \\ #4 & #5 & #6 \end{array}\right)}
\newcommand{\sixj}[6]{\left\{\begin{array}{ccc}#1 & #2 & #3 \\ #4 & #5 & #6 \end{array}\right\}}
\newcommand{\ninej}[9]{\left\{\begin{array}{ccc}#1 & #2 & #3 \\ #4 & #5 & #6 \\ #7 & #8 & #9 \end{array}\right\}}

\huge

\begin{center}
Total number of $J$ levels for identical particles in a single-$j$ shell using coefficients of fractional parentage
\end{center}

\vspace{0.5cm}

\large

\begin{center}
Jean-Christophe Pain\footnote{jean-christophe.pain@cea.fr}
\end{center}

\normalsize

\begin{center}
CEA, DAM, DIF, F-91297 Arpajon, France
\end{center}

\vspace{0.5cm}


\begin{abstract}
Analytical expressions of the total number of $J$ levels for three and four fermions in a nuclear $j^n$ shell are provided. The formulas were derived using a combination of sum rules for coefficients of fractional parentage, and ``unusual'' identities, \emph{i.e.} which do not contain the weighting factor $(2J+1)$ involving $6j$ and $9j$ symbols.
\end{abstract}

\section{\label{sec1} Introduction}

The single-$j$ shell is a model system and a cornerstone of nuclear-structure studies \cite{TALMI93,BAYMAN63,MCCULLEN64}. Many aspects of such a paradigm were investigated, like symmetries \cite{ZAMICK08}, two-body random Hamiltonians \cite{ZHAO01,MULHALL00} or isospin relations. For instance, Zhao and Arima showed that a system of three fermions is exactly solvable for a single-$j$ shell in the presence of an angular-momentum $J$-pairing interaction \cite{ZHAO04}. In a work on the number of pairs of nucleons with specific angular momentum, Zamick found simplified relations for a system of four nucleons with isospins zero and two \cite{ZAMICK07}, and Zamick and Escuderos investigated symmetries arising when two-body interaction matrix elements with isospin zero are set equal to a constant in a single-$j$ shell calculation \cite{ZAMICK12}. In Ref. \cite{FU13}, Fu \emph{et al.} studied the spin-$J$ nucleon pair approximation for a single-$j$ shell. In particular, they concluded that $J$-pair approximations are very good for $J\approx 2j$. The determination of the number of spin-$J$ states (or levels) for $n$ identical particles in a single-$j$ shell (denoted here $N(J,j,n)$ following the notation of Talmi's article \cite{TALMI05}), first addressed by Bethe \cite{BETHE36}, is a fundamental issue of nuclear-structure theory \cite{DESHALIT04}. The total number of states is equal to the degeneracy of the configuration, {\emph{i.e.}

\begin{equation}
\sum_J(2J+1)N(J,j,n)=\bin{2j+1}{n},
\end{equation}

\noindent where $\bin{k}{\ell}=k!/(\ell!(k-\ell)!)$ is the binomial coefficient. Many works were dedicated to the determination of an algebraic expression for $N\left(J,j,n\right)$. As an example, in a study of the quantum Hall effect \cite{GINOCCHIO93}, Ginocchio and Haxton obtained a simple formula for $N\left(0,j,4\right)$, which is also equal to $N\left(j,j,3\right)$. Zhao and Arima gave empirical formulas of $N(J,j,n)$ for three, four and five particles \cite{ZHAO03}. Zamick and Escuderos interpreted the Ginocchio-Haxton formula by a combinatorial approach for $J=j$ with $n$=3 \cite{ZAMICK05b} and Talmi derived a recursion relation for $N\left(J,j,n\right)$ of $n$ fermions in a $j$ orbit in terms of $k\le n$ fermions in a $(j-1)$ orbit \cite{TALMI05}. In Refs. \cite{ZHAO05c,ZAMICK05d}, the authors extended the studies for $n$=3 and $n$=4 to the number of states with given spin $J$ and isospin $T$. The number of states of a given spin was found to be closely related to the sum rules of many six-$j$ (denoted $6j$ throughout the paper) and nine-$j$ (denoted $9j$ in the following) symbols \cite{ZHAO05b,PAIN11b}, as well as to coefficients of fractional parentage \cite{ZAMICK07b}. Zamick and Escuderos found a relationship between coefficients of fractional parentage obtained from the principal-parent method and from a seniority classification \cite{ZAMICK06}. They applied it to Redmond's recursion relation formula \cite{REDMOND54}, transformed it to the seniority scheme and used it for the determination of the number of spin-$J$ states for $J=j$ (result previously obtained by Rowe and Rosensteel using a quasi-spin formulation \cite{ROWE01,ROSENSTEEL03}) and for $J=j+1$, result previously obtained by Zhao and Arima \cite{ZHAO03}. It is worth mentioning that Qi \emph {et al.} \cite{QI10} found an alternative proof of the Rowe-Rosensteel proposition, and that Wang and Xu showed that the partial conservation of seniority in $j$=9/2 shels is a consequence of special properties of some one-particle coefficients of fractional parentage \cite{WANG12}. Jiang \emph {et al.} derived analytical formulas for the number of spin-$J$ states for three identical particles, in a unified form for bosons and fermions \cite{JIANG13}. The idea consists in considering $\tilde{n}$ virtual bosons with spin 3/2, where $\tilde{n}=2j-2$ if one studies fermions in a single-$j$ shell, or $2\ell$ if one studies bosons with spin $J$. Three years ago, Bao \emph{et al.} derived recursive formulas by induction with respect to $n$ and $j$ and applied them to systems of two, three and five identical particles \cite{BAO16}. More recently, I published recursion relations obtained using the generating-function technique \cite{PAIN11}, as well as a study of the odd-even staggering (the number of odd-$J$ states is larger than the number of even-$J$ states) \cite{PAIN18}. However, the total number of levels

\begin{equation}
N_{\mathrm{tot}}(j,n)=\sum_J N(J,j,n).
\end{equation}

\noindent is not known analytically. Nevertheles, the knowledge of $N(J,j,n)$ and of such a quantity $N_{\mathrm{tot}}(j,n)$ is useful for nuclear-structure calculations, especially for the determination of the sizes of the Hamiltonian matrices. In the present work, I propose an exact expression for $N_{\mathrm{tot}}$ for 3 and 4 fermions. The expression is based on two basic properties: the first one is a sum rule involving coefficients of fractional parentage \cite{AYOUB77}, and the second one is an unusual sum rule (in the sense that it does not involve a weighting factor $(2J+1)$). For the three-particle case, I use an unusual sum rule of $6j$ coefficients \cite{VANAGAS61,KANCEREVICIUS90,PAIN11}, and in the four-particle case an unusual sum rule of $9j$ symbols derived by Elbaz using results obtained by Dunlap and Judd \cite{DUNLAP75,ELBAZ85}. Such unusual sum rules can not be found in the reference textbook of Varshalovich, Moskalev and Khersonskii \cite{VARSHALOVICH88}. All the results presented here can be generalized including isospin \cite{AYOUB77}.

In Sec. \ref{sec2}, the definition as well as some properties of fractional parentage coefficients are briefly recalled. The number of levels in a single-$j$ nuclear shell is derived in Sec. \ref{sec3} for three fermions (protons or neutrons), and in Sec. \ref{sec4} for four fermions.

\section{Coefficients of fractional parentage}\label{sec2}

Since their inception by Racah \cite{RACAH43}, the coefficients of fractional parentage are key ingredients of nuclear-structure calculations. The literature about that subject is very abundant, I just mention below a few works which are related to the present study. Bayman and Lande published tables of identical-particle coefficients of fractional parentage \cite{BAYMAN66}. Shlomo proposed to calculate one-particle $jj$-coupling coefficients of fractional parentage of both protons and neutrons using a recursion relation \cite{SHLOMO72}. In 1980, Mari\'c and Popovi\'c-Bo$\mathrm{\check{z}}$i\'c published coefficients of fractional parentage for an arbitrary number of $j=1$ bosons \cite{MARIC80}. Deveikis and Kamuntavi$\mathrm{\check{c}}$ius developed a new procedure for the calculation of coefficients of fractional parentage, free from the numerical diagonalization, orthogonalization and even group theoretical antisymmetric states classification \cite{DEVEIKIS95}. In 1997, Towner published tables of coefficients of fractional parentage for the $j$=7/2 shell in a seniority basis with good isospin \cite{TOWNER97}. The coefficients of fractional parentage are defined through

\begin{eqnarray}
\psi\left(j^n;J\gamma\right)=\sum_{J'\gamma'}\left[j^{n-1}\left(J'\gamma'\right)j\left.|\right\}j^nJ\gamma\right]\times\left[\psi\left(j^{n-1};J'\gamma'\right)\eta(j)\right]^J,
\end{eqnarray}

\noindent where $\psi\left(j^ n;J\gamma\right)$ stands for a normalized $n$-particle state totally antisymmetric in all $n$ particles, and $j$ is the angular momentum of each of the identical particles. The wave function is specified by the total angular momentum $J$ and the other quantum numbers, such as seniority, are symbolized by $\gamma$. $\eta(j)$ is a single-particle wave function and the states $\psi\left(j^{n-1};J'\gamma'\right)$ and $\eta(j)$ are coupled to angular momentum $J$. 

The coefficients of fractional parentage $\left[j^{n-1}\left(J'\gamma'\right)j\left.|\right\}j^nJ\gamma\right]$ satisfy the orthogonality relation

\begin{eqnarray}\label{normn}
\sum_{J'\gamma'}\left[j^{n-1}\left(J'\gamma'\right)j\left.|\right\}j^nJ_1\gamma_1\right]\left[j^{n-1}\left(J'\gamma'\right)j\left.|\right\}j^nJ_2\gamma_2\right]=\delta_{\gamma_1\gamma_2}\delta_{J_1J_2}.
\end{eqnarray}

\section{Total number of levels for the three-electron single-$j$ shell}\label{sec3}

The antisymmetric normalized three-body state can be built as a linear combination of the totally antisymmetric three-particle states:

\begin{equation}\label{dec3}
\phi(j^2(J_1)j;J)=\sum_{\gamma}x(J_1,J,\gamma)\psi(j^3;J\gamma),
\end{equation}

\noindent with

\begin{equation}\label{norm3}
\sum_{\gamma}x^2(J_1,J,\gamma)=1.
\end{equation}

Taking the scalar product of Eq. (\ref{dec3}) with $\psi(j^3;J\gamma)$, one gets

\begin{equation}
x(J_1,J,\gamma)=\langle\psi(j^3;J\gamma)^{\dag}\phi(j^2(J_1)jJ)\rangle.
\end{equation}

It can be proven \cite{AYOUB77} that

\begin{equation}
x(J_1,J,\gamma)=\sqrt{6}\left[j^2\left(J_1\right)j\left.|\right\}j^3J\gamma\right]\mathcal{N}_3\left(J_1,J\right)
\end{equation}

\noindent with

\begin{eqnarray}
\frac{1}{\mathcal{N}_3\left(J_1,J\right)^2}=\left[\vphantom{\sixj{J_1}{j}{J}{J_1}{j}{j}}1+(-1)^{J_1}\right]\left[1+2\left(2J_1+1\right)\sixj{J_1}{j}{J}{J_1}{j}{j}\right]
\end{eqnarray}

\noindent and one has

\begin{eqnarray}\label{resint}
\sum_{\gamma}\left[j^2\left(J_1\right)j\left.|\right\}j^3J\gamma\right]^2=\frac{1}{6}\left[1+(-1)^{J_1}\right]\left[\vphantom{\sixj{J_1}{j}{J}{J_1}{j}{j}}1+2(2J_1+1)\times\sixj{J_1}{j}{J}{J_1}{j}{j}\right].
\end{eqnarray}

The normalization condition (\ref{normn}) enables one to write, for three particles:

\begin{equation}\label{norm}
\sum_{J_1}\left[j^2\left(J_1\right)j\left.|\right\}j^3J\gamma\right]^2=1
\end{equation}

\noindent and, summing Eq. (\ref{norm}) over $\gamma$ (which labels the independent states), one obtains the number of independent states

\begin{equation}
N(J,j,3)=\sum_{\gamma,J_1}\left[j^2\left(J_1\right)j\left.\left.\right|\right\}j^3J\gamma\right]^2
\end{equation}

\noindent giving, using Eq. (\ref{resint}):

\begin{eqnarray}\label{test1}
N(J,j,3)=\frac{1}{6}\sum_{J_1=J_{1,\mathrm{min}}}^{J_{1,\mathrm{max}}}\left[1+(-1)^{J_1}\right]\left[1+2(2J_1+1)\sixj{J_1}{j}{J}{J_1}{j}{j}\right],
\end{eqnarray}

\noindent where $J_{1,\mathrm{min}}=|J-j|$ and $J_{1,\mathrm{max}}=\min(2j,J+j)$. Equation (\ref{test1}) can be put in the form

\begin{table}[ht]
\begin{center}
\begin{tabular}{cccccc}\hline\hline
$J$ / Shell & $(7/2)^3$ & $(9/2)^3$ & $(11/2)^3$ & $(13/2)^3$ & $(15/2)^3$ \\\hline
 3/2 & 1 & 1 & 1 & 0 & 0 \\
 5/2 & 1 & 1 & 1 & 1 & 1 \\ 
 7/2 & 1 & 1 & 1 & 1 & 1 \\ 
 9/2 & 1 & 2 & 2 & 1 & 1 \\ 
 11/2 & 1 & 1 & 2 & 2 & 2 \\
 13/2 & 0 & 1 & 1 & 2 & 2 \\
 15/2 & 1 & 1 & 2 & 2 & 2 \\
 17/2 & 0 & 1 & 1 & 2 & 3 \\
 19/2 & 0 & 0 & 1 & 2 & 2 \\
 21/2 & 0 & 1 & 1 & 1 & 2 \\
 23/2 & 0 & 0 & 1 & 2 & 2 \\
 25/2 & 0 & 0 & 0 & 1 & 2 \\
 27/2 & 0 & 0 & 1 & 1 & 1 \\
 29/2 & 0 & 0 & 0 & 1 & 2 \\
 31/2 & 0 & 0 & 0 & 1 & 1 \\
 33/2 & 0 & 0 & 0 & 0 & 1 \\
 35/2 & 0 & 0 & 0 & 1 & 1 \\
 37/2 & 0 & 0 & 0 & 0 & 1 \\
 39/2 & 0 & 0 & 0 & 0 & 0 \\
 41/2 & 0 & 0 & 0 & 0 & 1 \\\hline
$N_{\mathrm{tot}}(j,3)$ & 6 & 10 & 15 & 21 & 28 \\\hline\hline
\end{tabular}
\end{center}
\caption{Numbers of spin-$J$ states for different $j^3$ shells.}\label{tab1}
\end{table}

\begin{eqnarray}
N(J,j,3)&=&\frac{1}{6}\left(\sum_{J_1=J_{1,\mathrm{min}}}^{J_{1,\mathrm{max}}}1\right)+\frac{1}{6}\sum_{J_1=J_{1,\mathrm{min}}}^{J_{1,\mathrm{max}}}(-1)^{J_1}+\frac{1}{3}\sum_{J_1=J_{1,\mathrm{min}}}^{J_{1,\mathrm{max}}}(2J_1+1)\sixj{J_1}{j}{J}{J_1}{j}{j}\nonumber\\
& &+\frac{1}{3}\sum_{J_1=J_{1,\mathrm{min}}}^{J_{1,\mathrm{max}}}(-1)^{J_1}(2J_1+1)\sixj{J_1}{j}{J}{J_1}{j}{j}\nonumber\\
\end{eqnarray}

\noindent and our purpose is to calculate

\begin{equation}
N_{\mathrm{tot}}(j,3)=\sum_JN(J,j,3).
\end{equation}

We have

\begin{equation}
\sum_{J_1=J_{1,\mathrm{min}}}^{J_{1,\mathrm{max}}}1=\left(J_{1,\mathrm{max}}-J_{1,\mathrm{min}}+1\right)
\end{equation}

\noindent and

\begin{equation}
\sum_{J_1=J_{1,\mathrm{min}}}^{J_{1,\mathrm{max}}}(-1)^{J_1}=\frac{1}{2}\left[(-1)^{J_{1,\mathrm{min}}}+(-1)^{J_{1,\mathrm{max}}}\right].
\end{equation}

Therefore, one has to calculate the two partial sums

\begin{equation}\label{rel1}
\sum_{J_1}(2J_1+1)\sixj{j}{j}{J_1}{j}{J}{J_1}
\end{equation}

\noindent and

\begin{equation}\label{rel2}
\sum_{J_1}(-1)^{J_1}(2J_1+1)\sixj{j}{j}{J_1}{j}{J}{J_1}.
\end{equation}

The $6j$ coefficients of the preceding equations for $J=j$ follow the sum rule \cite{ZHAO03,ZHAO05b,GINOCCHIO93,TALMI93,ZHAO05}:

\begin{equation}\label{eqz1}
\frac{1}{3}\left(\frac{2j+1}{2}+2\sum_{\mathrm{even}\; J_1}(2J_1+1)\sixj{j}{j}{J_1}{j}{j}{J_1}\right)=\left\lfloor\frac{2j+3}{6}\right\rfloor,
\end{equation}

\noindent where $\lfloor x\rfloor$ represents the integer part of $x$ (largest integer not exceeding $x$). Similarly, for $J=j+1$, one has

\begin{equation}\label{eqz2}
\frac{1}{3}\left(\frac{2j-1}{2}-2\sum_{\mathrm{even}\; J_1}(2J_1+1)\sixj{j}{j}{J_1}{j}{j+1}{J_1}\right)=\left\lfloor\frac{j}{3}\right\rfloor.
\end{equation}

Such sum rules are required in order to determine the number of independent interactions in a given $j$ shell that conserve seniority \cite{TALMI93}. More general sum rules are given by Zhao and Arima in the appendix of Ref. \cite{ZHAO05b}. The relations (\ref{eqz1}) and (\ref{eqz2}) are, for a half-integer $j$, particular cases of the identity:
 
\begin{equation}\label{s1}
\sum_{\mathrm{even}\; J_1}(2J_1+1)\sixj{j}{j}{J_1}{j}{J}{J_1}=
\left\{\begin{array}{ll}
\frac{3}{2}\left\lfloor\frac{2J+3}{6}\right\rfloor-\frac{J}{2}-\frac{1}{4} & \mbox{if $J\leq j$}\\
\frac{3}{2}\left\lfloor\frac{3j-3-J}{6}\right\rfloor+\frac{3}{2}\Delta_{J,j}-\frac{1}{2}\left\lfloor\frac{3j+1-J}{2}\right\rfloor & \mbox{if $J\geq j$}
\end{array}\right.,
\end{equation}

\noindent where

\begin{equation}
\Delta_{I,j}=\left\{
\begin{array}{ll}
0 \;\;\;\;\;\mathrm{if}\;\;\;\;\; (3j-3-J) & \mbox{mod $6=1$}\\
1 & \mbox{otherwise.}
\end{array}\right.
\end{equation}

The same summation over odd values of $J_1$ leads to:

\begin{equation}\label{s2}
\begin{array}{ll}
\sum_{\mathrm{odd}\; J_1}(2J_1+1)\sixj{j}{j}{J_1}{j}{J}{J_1} & =\frac{J}{2}+\frac{1}{4}-\frac{3}{2}\left\lfloor\frac{2j+3}{6}\right\rfloor \\
 & =\left\{\begin{array}{ll}
-1 & \mbox{if $2j=3k$} \\
0 & \mbox{if $2j=3k+1$}\\
1 & \mbox{if $2j=3k+2$}.
\end{array}\right.\end{array}
\end{equation}
 
All the sum rules given in Ref. \cite{ZHAO05b}, including Eqs. (\ref{s1}) and (\ref{s2}), involve the weighting factor $(2J_1+1)$. I would like to mention that Eq. (\ref{s1}) and Eq. (\ref{s2}) can be combined to obtain the following sum rules:

\begin{eqnarray}
\sum_{J_1}(2J_1+1)\sixj{j}{j}{J_1}{j}{J}{J_1}=\left\{\begin{array}{l}
(-1)^{J+j}\times t\;\;\;\;\mathrm{if}\;\;\;\;j\geq J,\\
\frac{(-1)^{t+1}}{2}\left[1+(-1)^ {j+J}\right]\;\;\;\;\mathrm{if}\;\;\;\;j<J,
\end{array}\right.
\end{eqnarray}

\noindent and

\begin{equation}
\sum_{J_1}(-1)^{J_1}(2J_1+1)\sixj{j}{j}{J_1}{j}{J}{J_1}=1-(J-r)+3\left\lfloor\frac{J-r}{3}\right\rfloor,
\end{equation}

\noindent with $r=3(1-t)/2$ and $t=(1+(-1)^{2j})/2$. Such relations were also derived by Vanagas and Batarunas in their paper on the characters of the symmetric group SO(3) \cite{VANAGAS61,KANCEREVICIUS90}. The total number of levels for three fermions in a single-$j$ shell reads

\begin{eqnarray}
N_{\mathrm{tot}}(j,3)&=&\sum_J N(J,j,3)=\frac{1}{6}\sum_{J=J_{\mathrm{min}}}^{J_{\mathrm{max}}}\left(J_{1,\mathrm{max}}-J_{1,\mathrm{min}}+1\right)+\frac{1}{12}\sum_{J=J_{\mathrm{min}}}^{J_{\mathrm{max}}}\left[(-1)^{J_{1,\mathrm{min}}}+(-1)^{J_{1,\mathrm{max}}}\right]\nonumber\\
& &+\frac{t}{3}\sum_{J=J_{\mathrm{min}}}^j(-1)^{J+j}+\frac{(-1)^{t+1}}{6}\sum_{J=j+1}^{J_{\mathrm{min}}}\left[1+(-1)^{J+j}\right]+\frac{1}{3}\sum_{J=J_{\mathrm{min}}}^{J_{\mathrm{max}}}(1+r-J)\nonumber\\
& &+\left\lfloor\sum_{J=J_{\mathrm{min}}}^{J_{\mathrm{max}}}\left(\frac{J-r}{3}\right)\right\rfloor.
\end{eqnarray}

\noindent Taking into account the fact that $J_{\mathrm{max}}=3(j-1)$ and $J_{\mathrm{min}}=1/2$, we obtain

\begin{equation}
\sum_{J=J_{\mathrm{min}}}^{J=J_{\mathrm{max}}}\left(J_{1,\mathrm{max}}-J_{1,\mathrm{min}}+1\right)=-\frac{21}{4}+3j(j+1),
\end{equation}

\begin{eqnarray}
\sum_{J=J_{\mathrm{min}}}^{J=J_{\mathrm{max}}}\frac{\left[(-1)^{J_{1,\mathrm{min}}}+(-1)^{J_{1,\mathrm{max}}}\right]}{2}=\frac{\left[(-1)^{j+1/2}+(-1)^{j-1/2}\right]}{4}+\frac{3}{2}-j,
\end{eqnarray}

\begin{equation}
\sum_{J=J_{\mathrm{min}}}^j(-1)^{J+j}=0,
\end{equation}

\begin{equation}
\frac{(-1)^{t+1}}{2}\sum_{J=j+1}^{J_{\mathrm{max}}}\left[1+(-1)^{J+j}\right]=\frac{1}{2}(3-2j),
\end{equation}

\begin{equation}
\sum_{J=J_{\mathrm{min}}}^{J_{\mathrm{max}}}(1+r-J)=\frac{3(5-6j)(2j-5)}{8}
\end{equation}

\noindent and

\begin{equation}
\left\lfloor\sum_{J=J_{\mathrm{min}}}^{J_{\mathrm{max}}}\left(\frac{J-r}{3}\right)\right\rfloor=\frac{1}{8}(2j-5)(6j-5).
\end{equation}

Finally, the total number of levels for three fermions in a single-$j$ shell is given by the very simple and compact expression

\begin{equation}
N_{\mathrm{tot}}(j,3)=\frac{\left(4j^2-1\right)}{8}.
\end{equation}

\section{Case of four particles}\label{sec4}

In the case of four fermions, following the same procedure and using the expression

\begin{equation}
\phi(j^2(J_1)j^2(J_2);J)=\sum_{\gamma}x(J_1,J_2,J,\gamma)\psi(j^4;J\gamma),
\end{equation} 

Ayoub and Mavromatis obtained \cite{AYOUB77}:

\begin{eqnarray}
N(J,j,4)&=&\frac{1}{4!}\sum_{J_1,J_2}\left[1+(-1)^ {J_1}\right]\left[1+(-1)^{J_2}\right]\nonumber\\
& &\times\left[\vphantom{\ninej{j}{j}{J_1}{j}{j}{J_2}{J_2}{J_1}{J}}1+(-1)^{J}\delta_{J_1,J_2}-4\times(-1)^{J}(2J_1+1)(2J_2+1)\ninej{j}{j}{J_1}{j}{j}{J_2}{J_2}{J_1}{J}\right],\nonumber\\
& &
\end{eqnarray}

\noindent which is also equal to

\begin{eqnarray}
N(J,j,4)&=&\frac{1}{4!}\sum_{J_1,J_2}\left[1+(-1)^ {J_1}\right]\left[1+(-1)^{J_2}\right]\nonumber\\
& &\times\left[\vphantom{\ninej{j}{j}{J_1}{j}{j}{J_2}{J_2}{J_1}{J}}1+(-1)^{J}\delta_{J_1,J_2}-4(2J_1+1)(2J_2+1)\ninej{j}{j}{J_2}{j}{j}{J_1}{J_2}{J_1}{J}\right]\nonumber\\
& &
\end{eqnarray}

\noindent and our purpose is to calculate

\begin{equation}
N_{\mathrm{tot}}(j,4)=\sum_JN(J,j,4).
\end{equation}

The most difficult quantity to calculate is

\begin{eqnarray}
S(J,j,4)=\sum_{J_1,J_2}\left[1+(-1)^{J_1}\right]\left[1+(-1)^{J_2}\right](2J_1+1)(2J_2+1)\ninej{j}{j}{J_2}{j}{j}{J_1}{J_2}{J_1}{J}.\nonumber\\
& &
\end{eqnarray}

It it simple to check that

\begin{equation}
S(J,j,4)=4S_{\mathrm{ee}}(J,j,4),
\end{equation}

\noindent where

\begin{eqnarray}
S_{\mathrm{ee}}(J,j,4)=\sum_{J_1~\mathrm{even},J_2~\mathrm{even}}(2J_1+1)(2J_2+1)\ninej{j}{j}{J_2}{j}{j}{J_1}{J_2}{J_1}{J}.
\end{eqnarray}

Zhao and Arima found that, for $J\ge 2j$ \cite{ZHAO05b}:

\begin{eqnarray}
S_{\mathrm{ee}}(J,j,4)=\frac{1}{2}\left\lfloor\frac{4j-J}{2}\right\rfloor\times\left\lfloor\frac{4j-J+2}{2}\right\rfloor+(-1)^J\left\lfloor\frac{4j+2-J}{4}\right\rfloor-6D_J
\end{eqnarray}

\noindent where

\begin{eqnarray}
D_J=\left(\left\lfloor\frac{J_0}{6}\right\rfloor+1\right)\left(3\left\lfloor\frac{J_0}{6}\right\rfloor+J_0~\mathrm{mod}~6\right)+\delta_{J_0~\mathrm{mod}~6,0},
\end{eqnarray}

\noindent with

\begin{equation}
J_0=2j-\frac{15}{4}-\frac{J}{2}+\frac{3}{4}(-1)^J.
\end{equation}

Unfortunately, as stated by Zhao and Arima, it is more difficult to obtain a closed formula in the case where $J\leq 2j-1$. However, they found

\begin{equation}
S_{\mathrm{ee}}(J,j,4)=\left\{
\begin{array}{ll}
2m-2 & \;\;\;\;\mathrm{for}\;\;\;\;J=0,\\
0 & \;\;\;\;\mathrm{for}\;\;\;\;J=1,\\
4-2m & \;\;\;\;\mathrm{for}\;\;\;\;J=2,\\
2m & \;\;\;\;\mathrm{for}\;\;\;\;J=3,\\
2 & \;\;\;\;\mathrm{for}\;\;\;\;J=4,\\
4-2m & \;\;\;\;\mathrm{for}\;\;\;\;J=5,\\
2+2m & \;\;\;\;\mathrm{for}\;\;\;\;J=6,\\
4 & \;\;\;\;\mathrm{for}\;\;\;\;J=7,\\
6-2m & \;\;\;\;\mathrm{for}\;\;\;\;J=8,\\
2+2m & \;\;\;\;\mathrm{for}\;\;\;\;J=9,\\
6 & \;\;\;\;\mathrm{for}\;\;\;\;J=10,\\
8-2m & \;\;\;\;\mathrm{for}\;\;\;\;J=11,\\
\vdots
\end{array}
\right.
\end{equation}

\noindent where $m=\left(j-\frac{3}{2}\right)~\mathrm{mod}~3$. They also noted the modular property

\begin{equation}
S_{\mathrm{ee}}(J,j,4)=S_{\mathrm{ee}}\left(J~\mathrm{mod}~12,j,4\right)+6\left\lfloor\frac{J}{12}\right\rfloor.
\end{equation}

In the present work, I decided to compute directly $N_{\mathrm{tot}}(j,4)$, by inverting the summations over $\left\{J_1,J_2\right\}$ and over $J$:

\begin{equation}
\sum_{J=J_{\mathrm{min}}}^{J_{\mathrm{max}}}\sum_{J_1=|J_2-J|}^{\min\left(J_2+J,2j\right)}\sum_{J_2=0}^{2j}\cdots=\sum_{J_1=0}^{2j}\sum_{J_2=0}^{2j}\sum_{J=|J_1-J_2|}^{J_1+J_2}\cdots,
\end{equation} 

\noindent with $J_{\mathrm{min}}=0$ and $J_{\mathrm{max}}=2(2j-3)$. One has

\begin{eqnarray}
N_{\mathrm{tot}}(j,4)&=&\sum_J N(J,j,4)=\frac{1}{4!}\sum_{J_1=0}^{2j}\sum_{J_2=0}^{2j}\left[1+(-1)^{J_1}\right]\left[1+(-1)^{J_2}\right]\nonumber\\
& &\times\left[\vphantom{\ninej{j}{j}{J_1}{j}{j}{J_2}{J_2}{J_1}{J}}\left(\sum_{J=|J_1-J_2|}^{J_1+J_2}1\right)+\delta_{J_1,J_2}\sum_{J=|J_1-J_2|}^{J_1+J_2}(-1)^J\right.\nonumber\\
& &\left.-4(2J_1+1)(2J_2+1)\sum_{J=|J_1-J_2|}^{J_1+J_2}\ninej{j}{j}{J_2}{j}{j}{J_1}{J_2}{J_1}{J}\right],
\end{eqnarray}

\noindent or

\begin{eqnarray}
N_{\mathrm{tot}}(j,4)&=&\frac{1}{4!}\sum_{J_1=0}^{2j}\sum_{J_2=0}^{2j}\left[1+(-1)^{J_1}\right]\left[1+(-1)^{J_2}\right]\left[\vphantom{\ninej{j}{j}{J_1}{j}{j}{J_2}{J_2}{J_1}{J}}\left(J_1+J_2-|J_1-J_2|+1\right)\right.\nonumber\\
& &+\left.\frac{\delta_{J_1,J_2}}{2}\left[(-1)^{|J_1-J_2|}+(-1)^{J_1+J_2}\right]-4(2J_1+1)(2J_2+1)\sum_{J=|J_1-J_2|}^{J_1+J_2}\ninej{j}{j}{J_2}{j}{j}{J_1}{J_2}{J_1}{J}\right].\nonumber\\
& &
\end{eqnarray}

Taking into account the fact that

\begin{eqnarray}
& &\frac{1}{4!}\sum_{J_1=0}^{2j}\sum_{J_2=0}^{2j}\left[1+(-1)^{J_1}\right]\left[1+(-1)^{J_2}\right]\left(J_1+J_2-|J_1-J_2|+1\right)=\frac{1}{72}(2j+1)\left(8j^2+2j+3\right)\nonumber\\
& &
\end{eqnarray}

\noindent and that

\begin{eqnarray}
& &\frac{1}{4!}\sum_{J_1=0}^{2j}\sum_{J_2=0}^{2j}\left[1+(-1)^{J_1}\right]\left[1+(-1)^{J_2}\right]\frac{\delta_{J_1,J_2}}{2}\left[(-1)^{|J_1-J_2|}+(-1)^{J_1+J_2}\right]=\frac{1}{12}(2j+1),
\end{eqnarray}

\noindent one has

\begin{eqnarray}
N_{\mathrm{tot}}(j,4)&=&\frac{1}{72}(2j+1)\left(8j^2+2j+3\right)+\frac{1}{12}(2j+1)-\frac{1}{3!}\sum_{J_1=0}^{2j}\sum_{J_2=0}^{2j}\left[1+(-1)^{J_1}\right]\left[1+(-1)^{J_2}\right]\nonumber\\
& &\times(2J_1+1)(2J_2+1)\sum_{J=|J_1-J_2|}^{J_1+J_2}\ninej{j}{j}{J_2}{j}{j}{J_1}{J_2}{J_1}{J}.\nonumber\\
& &
\end{eqnarray}

\begin{table}[ht]
\begin{center}
\begin{tabular}{cccccc}\hline\hline
$J$ / Shell & $(7/2)^4$ & $(9/2)^4$ & $(11/2)^4$ & $(13/2)^4$ & $(15/2)^4$ \\\hline
 0 & 1 & 2 & 2 & 2 & 3 \\
 1 & 0 & 0 & 0 & 0 & 0 \\ 
 2 & 2 & 2 & 3 & 4 & 4 \\ 
 3 & 0 & 1 & 1 & 1 & 2 \\ 
 4 & 2 & 3 & 4 & 5 & 6 \\
 5 & 1 & 1 & 2 & 3 & 3 \\
 6 & 1 & 3 & 4 & 5 & 7 \\
 7 & 0 & 1 & 2 & 3 & 4 \\
 8 & 1 & 2 & 4 & 6 & 7 \\
 9 & 0 & 1 & 2 & 3 & 5 \\
10 & 0 & 1 & 3 & 5 & 7 \\
11 & 0 & 0 & 1 & 3 & 4 \\
12 & 0 & 1 & 2 & 4 & 7 \\
13 & 0 & 1 & 0 & 2 & 4 \\
14 & 0 & 1 & 0 & 3 & 5 \\
15 & 0 & 0 & 0 & 1 & 3 \\
16 & 0 & 1 & 0 & 2 & 4 \\
17 & 0 & 0 & 0 & 1 & 2 \\
18 & 0 & 0 & 0 & 1 & 3 \\
19 & 0 & 0 & 0 & 0 & 1 \\
20 & 0 & 0 & 0 & 1 & 2 \\
21 & 0 & 0 & 0 & 0 & 1 \\
22 & 0 & 0 & 0 & 0 & 1 \\
23 & 0 & 0 & 0 & 0 & 0 \\
24 & 0 & 0 & 0 & 0 & 1 \\\hline
Total & 8 & 21 & 30 & 55 & 86 \\\hline\hline
\end{tabular}
\end{center}
\caption{Numbers of spin-$J$ states for different $j^4$ shells.}\label{tab2}
\end{table}
         
Therefore, in order to find $N_{\mathrm{tot}}$ in the case of four fermions, one has to calculate

\begin{equation}
\sum_J\ninej{j}{j}{J_2}{j}{j}{J_1}{J_2}{J_1}{J}.
\end{equation}

\noindent In section IV (entitled ``The unusual $j$-summation rule'') of his paper ``Usual and unusual summation rules over $j$ angular momentum'', Elbaz gives three summation rules pp. 730 and 731 \cite{ELBAZ85}: the first one is

\begin{equation}
\sum_J(-1)^{j_1+j_2+k+j}\sixj{j_1}{j_2}{J}{j_2}{j_1}{k}=D(j_1,j_2;k), 
\end{equation}

\noindent where $D$ is the coefficient introduced by Dunlap and Judd \cite{DUNLAP75}:

\begin{equation}\label{dun}
D_{J_a,J_b;k}=\frac{1}{2k+1}\left[\frac{\left(2J_a-k\right)!\left(2J_b+k+1\right)!}{\left(2J_b-k\right)!\left(2J_a+k+1\right)!}\right]^{1/2},
\end{equation}

\noindent the second unusual summation rule is 

\begin{equation}
\sum_{m_1}\threej{j_1}{m_1}{j_2}{M-m_1}{k}{-M}^2=\frac{\left\{j_1,j_2,k\right\}}{(2k+1)}, 
\end{equation}

\noindent where $\left\{j_1,j_2,k\right\}$=1 if $j_1$, $j_2$ and $k$ satisfy triangular relations and 0 otherwise, and finally the third summation rule is

\begin{eqnarray}
\sum_J\ninej{j}{j}{J_2}{j}{j}{J_1}{J_2}{J_1}{J}=\sum_{k=0}^{\min(2j,2J_1,2J_2)}(2k+1)(-1)^{\phi}D_{J_M,J_m;k}\sixj{j}{j}{k}{J_2}{J_2}{j}\sixj{j}{j}{k}{J_1}{J_1}{j},\nonumber\\
& &
\end{eqnarray}

\noindent with $\phi=J_1+J_2+k$, $J_m=\min\left(J_1,J_2\right)$, $J_M=\max\left(J_1,J_2\right)$. The latter sum is the one we need here, and finally, one has

\begin{eqnarray}
N_{\mathrm{tot}}(j,4)&=&\frac{2j+1}{72}\left[2j(4j+1)+9\right]-\frac{1}{3!}\sum_{J_1,J_2=0}^{2j}\left[1+(-1)^{J_1}\right]\left[1+(-1)^{J_2}\right]\nonumber\\
& &\times(2J_1+1)(2J_2+1)\sum_{k=0}^{\min(2j,2J_1,2J_2)}(2k+1)(-1)^{\phi}D_{J_M,J_m;k}\nonumber\\
& &\left.\times\sixj{j}{j}{k}{J_2}{J_2}{j}\sixj{j}{j}{k}{J_1}{J_1}{j}\right],\nonumber\\
\end{eqnarray}

\noindent which is the most compact expression I could obtain.

\vspace{0.5cm}

It is finally worth mentioning that, knowing the difference between odd and even-$J$ states (see my previous work on the odd-even staggering in a single-$j$ shell \cite{PAIN18}):

\begin{equation}
N_{\mathrm{even}}(j,4)-N_{\mathrm{odd}}(j,4)=\bin{2j+1}{2}
\end{equation}

\noindent and 

\begin{equation}
N_{\mathrm{odd}}(j,4)+N_{\mathrm{even}}(j,4)=N_{\mathrm{tot}}(j,4),
\end{equation}

\noindent one gets

\begin{equation}
\left\{\begin{array}{l}
N_{\mathrm{even}}=\frac{1}{2}\left[N_{\mathrm{tot}}(j,4)+\frac{(2j-1)(2j+1)}{2}\right]\\
N_{\mathrm{odd}}=\frac{1}{2}\left[N_{\mathrm{tot}}(j,4)-\frac{(2j-1)(2j+1)}{2}\right]\\
\end{array}\right..
\end{equation}

\section{Conclusion}

I have presented the expression of the total number of spin states (or levels) for 3 and 4 particles in a single-$j$ shell. The calculation relies on sum rules involving coefficients of fractional parentage, and unusual (or ``anomalous'') identities for $6j$ and $9j$ symbols. The latter relations are unusual in the sense that they do not contain the prefactor $(2J+1)$, $J$ being the angular momentum over which the summation is performed. The procedure used in the present work can in principle be generalized to any number of particles. However, as the number of particles increases, the normalization coefficients (equivalent to $\mathcal{N}_3$ in the three-fermion case) become more tedious to determine. This is also the case of the unusual sum rules, which are not known at the moment, and which derivation, even with the help of graphical methods, should be lenghty. Indeed, the calculations involve $3(n-1)j$ symbols ($12j$ symbols for $n=5$, $15j$ symbols for $n=6$, etc.).

\end{document}